\documentclass[a4paper,11pt]{article}
\pdfoutput=1 % if your are submitting a pdflatex (i.e. if you have images in pdf, png or jpg format)

\usepackage{jinstpub} % for details on the use of the package, please see the JINST-author-manual
\usepackage{soul} % for strike text

\usepackage{lineno}
\title{\boldmath A metamaterial with applications in broad band antennas used in radio astronomy and satellite communications}

\author[a,b,1]{J.~De~Miguel,\note{Corresponding author (J.~De~Miguel is now with Riken)}}
\author[c]{C.~Franceschet,}
\author[c]{S.~Realini,}
\author[a,b]{and P.~Fuerte-Rodríguez}
\affiliation[a]{Instituto de Astrof\'isica de Canarias, E-38200 La Laguna, Tenerife, Spain}
\affiliation[b]{Departamento de Astrof\'isica, Universidad de La Laguna, E-38206 La Laguna, Tenerife, Spain}
\affiliation[c]{Universit\'a degli Studi di Milano, Via Celoria 16, 20133 Milano, Italy}
\emailAdd{javier.miguelhernandez@riken.jp}

\abstract{Electromagnetic metamaterials at microwave frequencies are well established in industry and research. Recent work has shown how a specific kind of metallic metamaterial can contribute towards improving the performance of the feedhorn antennas used in radio astronomy and satellite telecommunications. In this article, we discuss an innovative type of meta-ring of remarkable manufacturability, able to improve the state of the art in these fields. A pioneering meta-horn antenna formed of meta-rings is then fabricated and characterized in the laboratory. It shows an excellent feature on an octave bandwidth, especially in terms of cross-polarization, a key figure of merit in both radio astronomy and telecommunications, and also side-lobe level, return-loss and gain.}


\arxivnumber{2108.05648}

\begin{document}
\maketitle
\flushbottom

%----------------------------------------
%% main text
\section{Introduction}
\label{Intro}

Electromagnetic metamaterials at microwave frequencies are well established in both industry and research (e.g., \cite{Schurig}, \cite{Sievenpiper}, \cite{Sievenpiper2} and \cite{Sievenpiper3}). They are distinguished by their electromagnetic properties, which differ from those given by their basic constituent as the consequence of a smart structure. Recent research led us to think that electromagnetic metamaterials can also be used in the manufacture of antennas for satellite communications \cite{Lier}~\cite{ADA}~\cite{Miguel_Hern_ndez_2019}~\cite{6493418}~\cite{8752466}. Horn-antennas made of metamaterials, or meta-horns, could give a significant improvement to the state of the art in the field of feedhorn antennas with low cross-polarization and sidelobe levels, a technology that has been static for several decades owing to insurmountable limitations in the bandwidth of corrugated feedhorns \cite{Clarricoats}. Unfortunately, the fabrication of metamaterial structures is challenging due to their geometric complexity and (frequency-dependent) reduced size required for applications in satellite communications and astronomy in the microwave range. This is probably the reason why this technology is not fully implanted nowadays in either radio astronomy or telecommunications.

In \cite{Miguel_Hern_ndez_2019} we theorized on cross-polarization of feed-horn antennas and we presented the topology of a new metamaterial able to surpass some limitations of corrugated horns. In \cite{DEMIGUELHERNANDEZ2019103195} we explored the manufacturability of such a metamaterial using different techniques. In this work, we present a prototype of a meta-horn made of this metamaterial, which shows promising results. Success in this research could provide important benefits for astronomy, since the sensitivity of broad band receivers scales as $\sqrt{\Delta \nu }$, i.e., the square root of bandwidth obtained in comparison with a classical corrugated horn-antenna \cite{Dicke1946TheMO}. Meta-horns could provide excellent performance on very wide frequency ranges not reachable with classic corrugated horn-antennas. Thanks to the ability to span multiple bands typically covered by several devices, they could relax the complexity of wide-band receivers in terms of hardware and associated costs. Moreover, current satellites tend to transmit through the K$_a$ and K$_u$ bands, which are too wide to be covered with a standard corrugated horn antenna. The novel technique presented in this work can cover both bands simultaneously with a single receiver in the frequency range 10--30~GHz. This corresponds to $B_f=3$, where $B_f$ is the bandwidth factor, i.e. the ratio between the maximal and minimal frequency in a band. The theoretical limit for the meta-horn presented here is around $B_f\sim5$; however, we aimed at a lower value to make our objective easier to achieve. Note that this is still a good improvement if compared to standard antennas, which typically yield $B_f\simeq 1.4$, commonly referred as a 40\% or 1.4:1 \cite{Clarricoats}.

This article is structured as follows. In Sec.~\ref{TheoryFundamentals} we introduce the fundamentals of reflector antennas and metamaterials relevant for this article. Section~\ref{Fabrication} presents a novel feedhorn antenna made of a new class of metamaterial, or meta-ring, and its experimental demonstration. Finally, in Sec.~\ref{Results} conclusions and future directions are discussed.

\section{Theoretical fundamentals}
\label{TheoryFundamentals}
The field generated at the optical focus of a paraboloid with large \textit{focal length} can be divided into two polarizations: a P-polarization, or also called transverse magnetic (TM) component, polarized in the plane of incidence, with its electric (E) component oriented normally to the plane on which the beam is projected, and a S-polarization, or transverse electric (TE) pattern, polarized normally to the plane of incidence \cite{Minnet}. The feedhorn which transmits both the azimuthal and longitudinal components with null cross-polarization\footnote{Cross-polarization could be qualitatively understood as crosstalk between the TE and TM modes through the antenna reducing the purity of the transmitted signal~\cite{1140406}.} must have inner surfaces satisfying the following expression:

\begin{equation}
Z^{S}Z^{P}=-Z_0^2 \;,
\label{equation_1}
\end{equation}
where $Z^{S,P}$ is the surface impedance for both polarizations (or, equivalently, spatial directions from the surface) and $Z_0$ is the impedance of the vacuum. The relation in (\ref{equation_1}) is known as the {\it balanced hybrid-mode condition} and guarantees that the HE$_{11}$ mode is \textit{fundamental} and of a high cross-polarization purity.
In \cite{Minnet} it was suggested that a radially corrugated cylindrical waveguide can be used in order to support the hybrid-mode condition over a limited band. Other authors give a more in-depth treatment of the corrugated feedhorn technology \cite{Clarricoats}, which efficiently exploits modes hybridization to push the bandwidth factor to around 1.4:1.

\begin{figure}
    \includegraphics[width=0.75\textwidth]{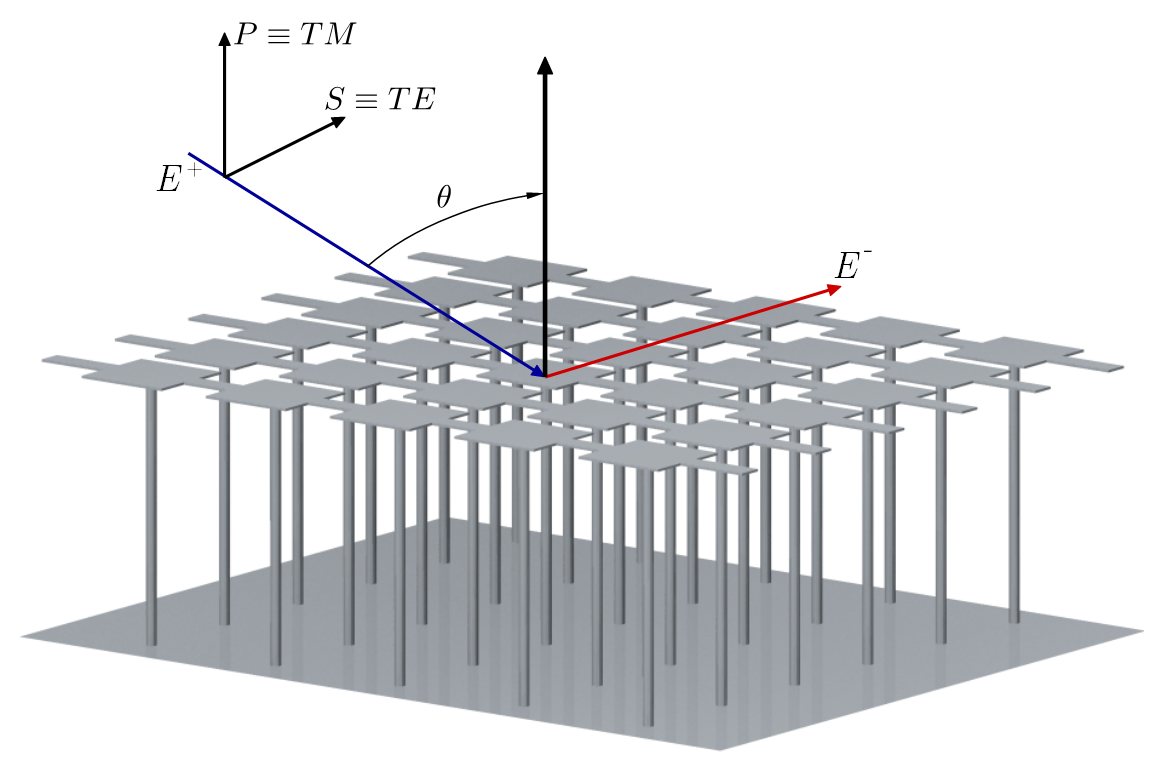}
    \centering 
    \caption{Schematic of $S$- and $P$-polarization of electromagnetic modes. Incoming waves (blue) are reflected (red) on a metamaterial surface, allowing one to estimate the polarization dependent surface impedance. The meta-surface shown in the figure is arbitrarily based on a design presented in \cite{ADA}.}
    \label{fig_1}
\end{figure}

To obtain an even wider bandwidth, 2:1 and beyond, while preserving low cross-polarization levels, we designed a new metamaterial with a surface impedance $Z^{S,P}$ equal to the critical value, similar to \cite{Lier}. It is well known that the reflection coefficient for a wave travelling through vacuum and perpendicularly to a plane surface is

\begin{equation}
\Gamma=\frac{E^-}{E^+}=\frac{Z_s-Z_0}{Z_s+Z_0} \;,
\label{equation_2}
\end{equation}
where $E^+$ represents the forward incident wave and $E^-$ represents the backward reflected wave, $Z_s$ is the impedance of the (plane) surface and $Z_0$ is the impedance of the vacuum. This is represented in Fig. \ref{fig_1}. 

Therefore, it is possible to obtain expressions for TE and TM modes being reflected by a surface with characteristic $Z^{S,P}$ impedance. The effective free-space impedance in this geometry is a function of the incidence angle $\theta$, which is arbitrary. For $P$-polarized modes

\begin{equation}
TM\equiv P
\begin{cases} 
 |H\big|= H_0 \\
 |E\big|=E_0\mathrm{cos}\theta
\end{cases} 
\longrightarrow Z_0^{P}(\theta)=\frac{E^{P}(\theta)}{H^{P}(\theta)}=Z_0\mathrm{cos}\theta \;,
\label{equation_3}
\end{equation}
and for $S$-polarized modes

\begin{equation}
TE\equiv S
\begin{cases} 
 |H\big|= H_0 \mathrm{cos}\theta\\
 |E\big|=E_0
\end{cases} 
\longrightarrow Z_0^{S}(\theta)=\frac{E^{S}(\theta)}{H^{S}(\theta)}=\frac{Z_0}{\mathrm{cos}\theta} \;.
\label{equation_4}
\end{equation}

From equations (\ref{equation_2}), (\ref{equation_3}) and (\ref{equation_4}), we can derive the expression of the surface impedance for the reflected $P$-modes

\begin{equation}
Z_s^{P}=Z_0 \, \mathrm{cos} \theta \, \frac{1+\Gamma^{P}}{1-\Gamma^{P}}\;,
\label{equation_6}
\end{equation}
and the reflected $S$-modes

\begin{equation}
Z_s^{S}=\frac{Z_0}{\mathrm{cos} \theta} \frac{1+\Gamma^{S}}{1-\Gamma^{S}}\;.
\label{equation_5}
\end{equation}

Finite element method (FEM) 3D electromagnetic simulations\footnote{Running on CST Studio Suite or similar softwares.} allow one to estimate the characteristic impedance $Z^{S,P}$ of each mode, once the reflection coefficients $\Gamma^{S,P}$ have been calculated. The $Z^{S,P}$ values are plugged in (\ref{equation_1}) to verify if the meta-surface fulfils the hybrid-mode condition.

In the simulations, a plane wave is reflected on a metamaterial containing an E-field probe over its surface. By comparing the E-field probed in the same position with and without the presence of the metamaterial under test, we can subtract the incoming and reflected signals to obtain the reflection coefficient and surface impedance for each $S$ or $P$ polarization from equations (\ref{equation_5}) and (\ref{equation_6}) (more details are given in \cite{Lier}, \cite{ADA}, \cite{Miguel_Hern_ndez_2019}, \cite{DEMIGUELHERNANDEZ2019103195}).

\begin{figure}
\centering
    \includegraphics[width=.3\textwidth]{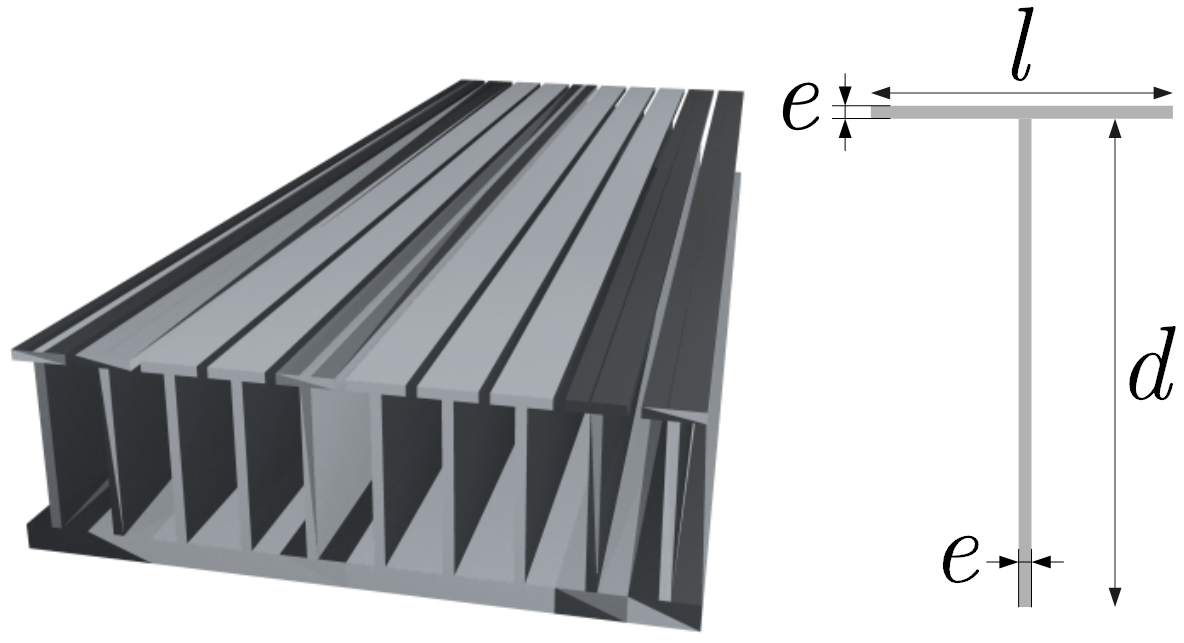}\hfill
    \includegraphics[width=.3\textwidth]{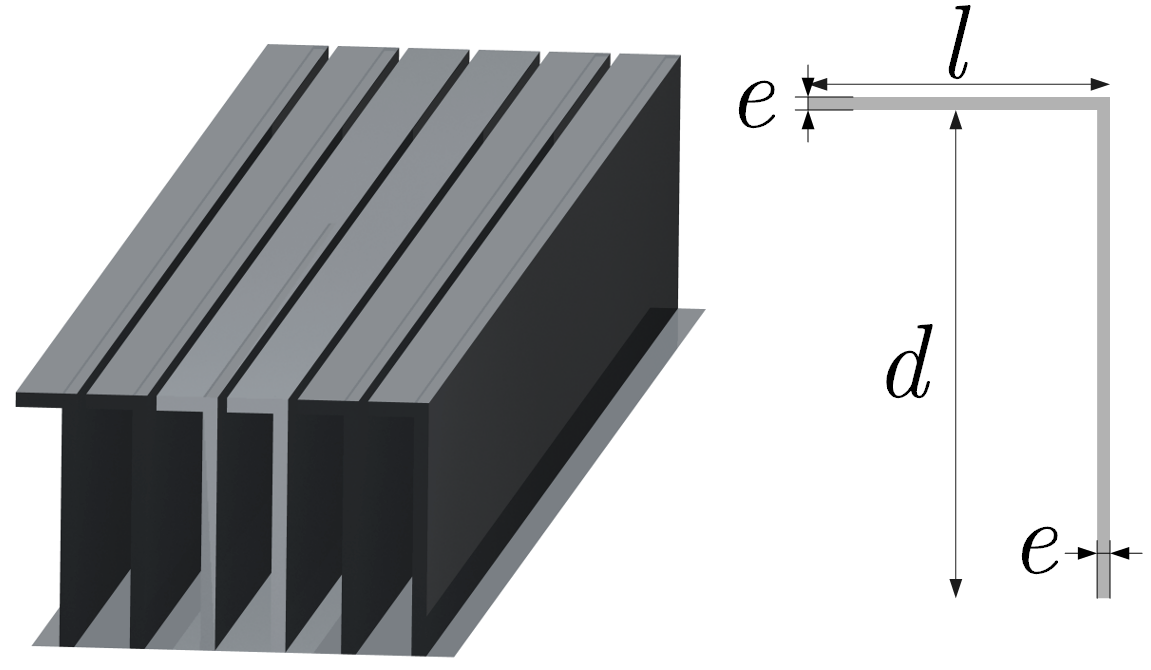}\hfill
    \includegraphics[width=.3\textwidth]{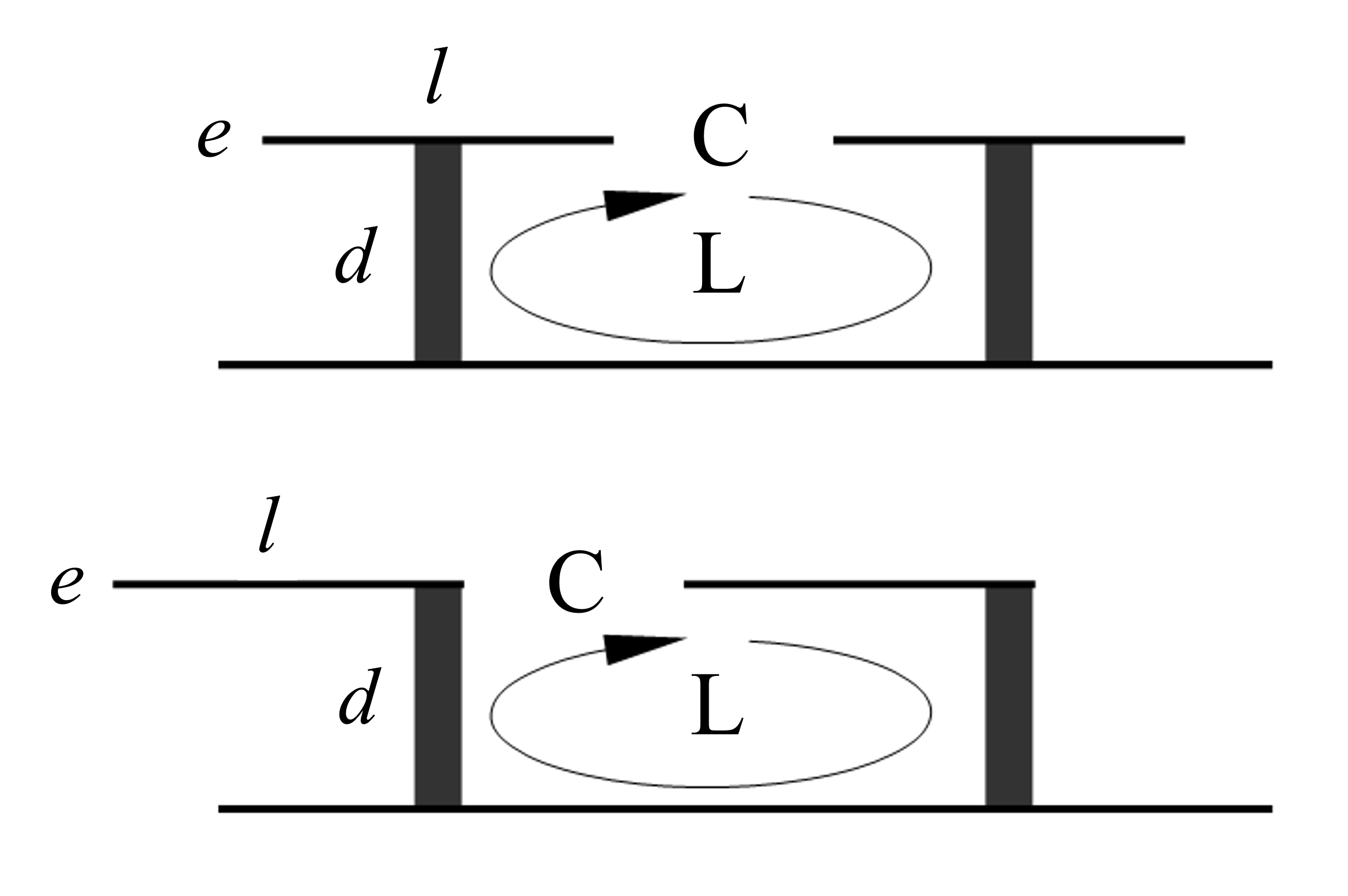}
    \caption{Simplified configurations of the top-plated (\textit{left}) and ``inverted L'' (\textit{center}) corrugations of the metamaterial studied in \cite{DEMIGUELHERNANDEZ2019103195}, where $e=0.4$~mm, $l=1.66$~mm, $d=6.5$~mm and the free space between adjacent teeth of the corrugations is 0.25~mm. \textit{Right}: transverse section of the top-plated (top) and ``inverted L'' (bottom) topologies. In an equivalent LC-model, the surface impedance of the metamaterial is given by $Z_s=\frac{j\omega \mathrm{L}}{1-\omega^2\mathrm{LC}}$, where L and C are inductance and capacitance, respectively~\cite{Sievenpiper}. The inductance is a result of the circulation of the wave around the two parallel plates which are shorted at one end, while the capacitive part of the surface reactance is a result of the wave passing through two parallel plates.}
    \label{fig_2}
\end{figure}

\begin{figure}
\centering 
    \includegraphics[trim=15 45 15 30, clip, width=0.55\textwidth]{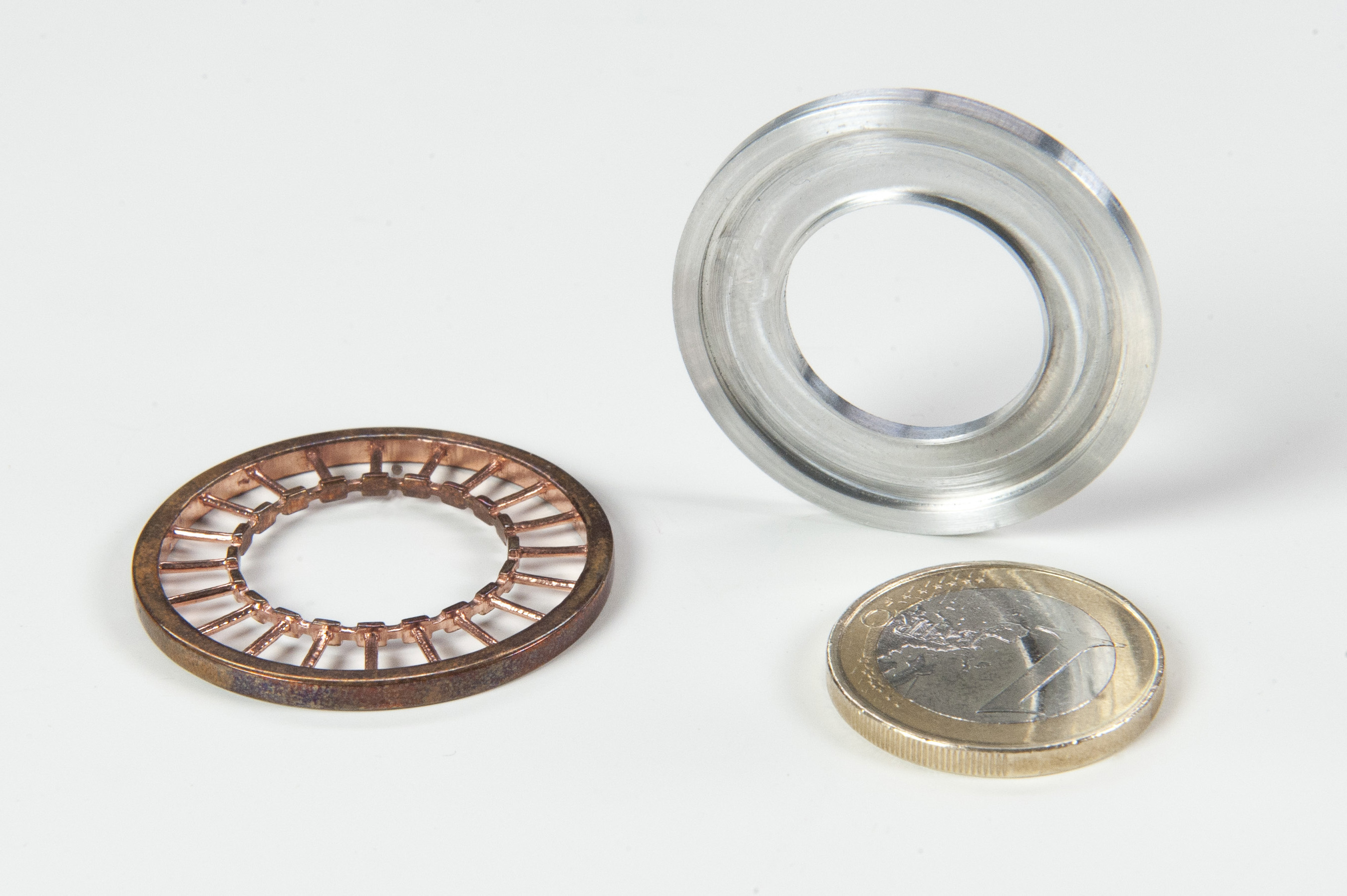}
    \caption{On the left, a copper meta-ring based on the structure shown in Fig.~\ref{fig_1}, fabricated by additive printing; the aluminum meta-ring on the right is based on the structure shown in Fig.~\ref{fig_2} and it is manufactured by traditional machining at a lower cost and with a higher precision~\cite{DEMIGUELHERNANDEZ2019103195}.}
    \label{fig_8}
\end{figure}

Left and center panels of Fig.~\ref{fig_2} show two simple structures of a novel 3D metamaterial, which are equivalent from an electromagnetic perspective, as explained in the right panel of the same figure. Their concept relies on a simplification of the structure shown in Fig.~\ref{fig_1}, which allows for an easy manufacturability. The simplified design not only guarantees performance comparable to that obtained for the model in Fig.~\ref{fig_1}, but also extends it to an even wider operating bandwidth, as discussed in \cite{DEMIGUELHERNANDEZ2019103195}. It is convenient to remark that the aim of these simulations is to design the inner surface of a horn-antenna, made up of tens or hundreds of rings based on the design in Fig.~\ref{fig_2} joined together. Fig.~\ref{fig_8} shows a first sample of a meta-ring. The technique to manufacture the meta-horn will be adequately discussed later in this work, and a proof-of-concept prototype will be presented.

\begin{figure}
    \includegraphics[width=0.5\textwidth]{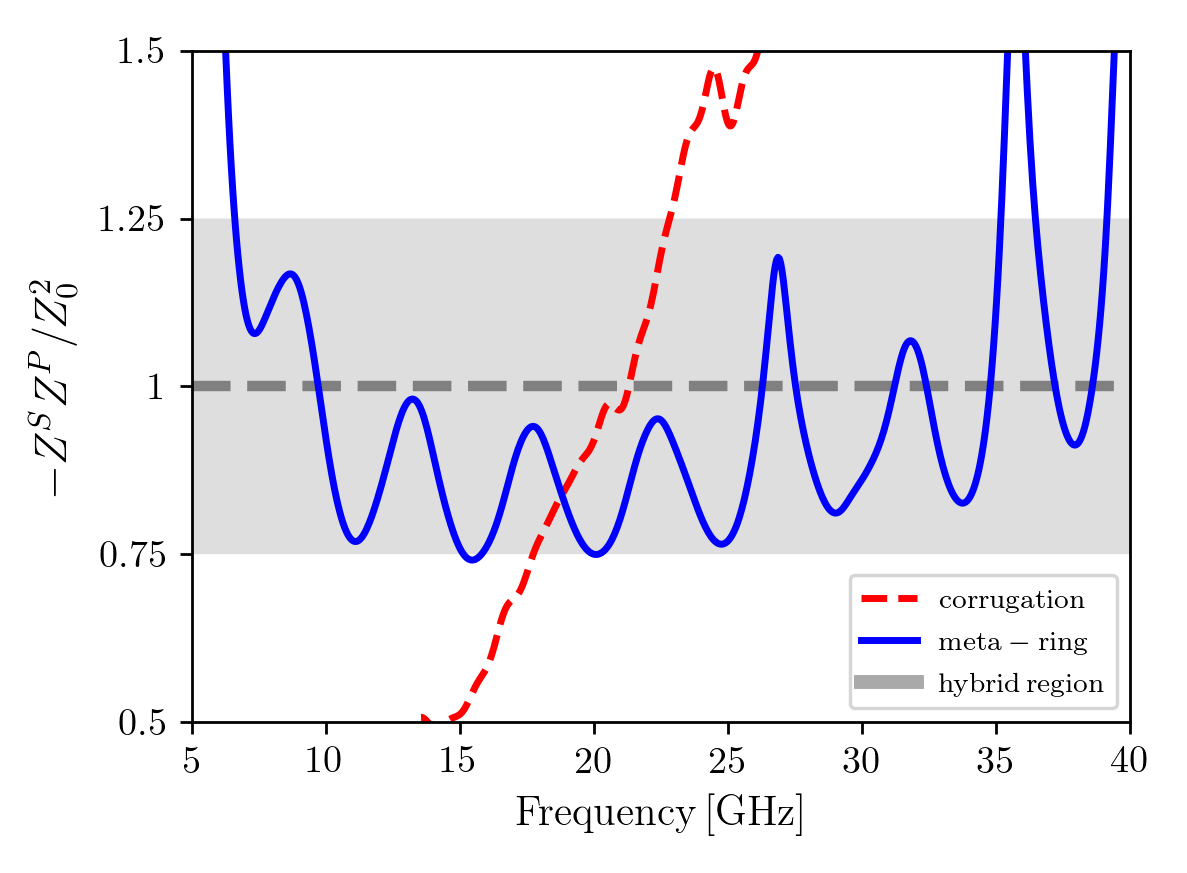}
    \centering 
    \caption{The blue line shows the metamaterial FEM-simulations for both top-plated and ``inverted L'' structures, since they present identical results. As a reference, the dashed red line represents an equivalent model using corrugations instead of metamaterials. Confirming previous expectations, the bandwidth of the corrugation is limited to a few tens of percent, while the metamaterial allows impedance tuning resulting in a wider band.}
    \label{fig_5}
\end{figure}

\begin{figure}[ht!]
    \includegraphics[width=.7\textwidth]{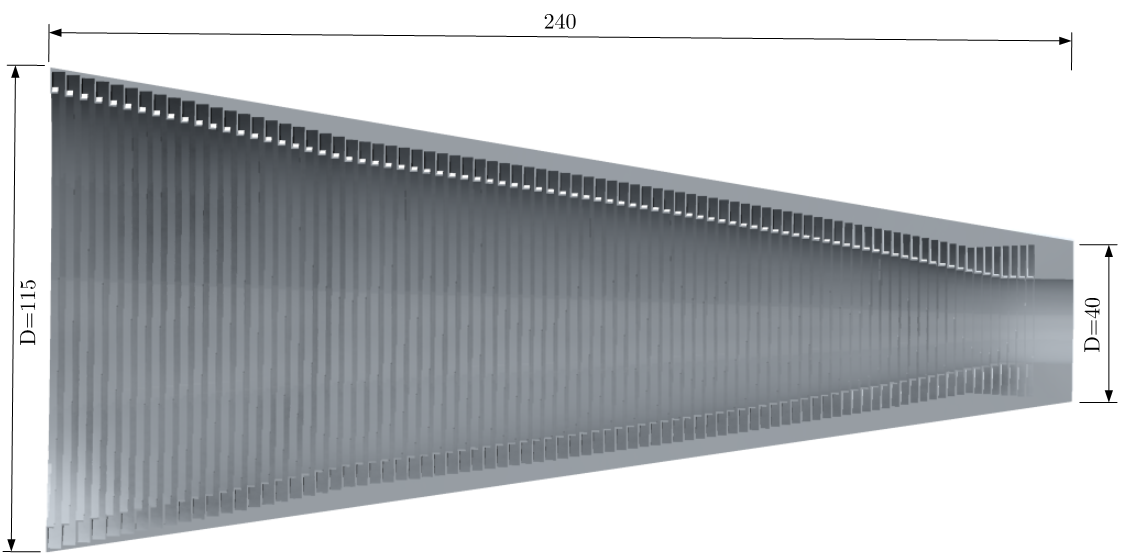}
    \centering 
    \begin{small}
        \put(-18,95){\vector(0,-1){20}}
        \put(-33,98){waveguide}
        \put(-39,110){\vector(0,-1){25}}
        \put(-53,118){mode}
        \put(-60,112){converter
        \put(-145,13){\vector(0,-1){18}}
        \put(-145,13){\vector(-4,-1){40}}
        \put(-175,15){conical profiles}
        %\put(-250,15){\vector(-2,-1){10}}
        \put(-252,20){aperture}
        }
    \end{small}
    \caption{Design of meta-horn (gross dimensions in millimeters). The methods used are the usual ones in corrugated horns \cite{Clarricoats, 1173026}. The dimensions of the TE$_{11}$-HE$_{11}$ converter vary from about $\lambda/2$ at the feeding waveguide to $\sim\!\!\lambda/4$ depth. Two different taper profiles differ by a few degrees of inclination on their path to the aperture. The longitudinal parametrization of the profiling of the `teeth' ($l$ and $d$ in Fig.~\ref{fig_2}) was aided by computational optimization aimed at simultaneously reducing side-lobes and cross-polarization effects, while its thickness is always set around $e\!\gtrsim \!0.4$~mm, coinciding with the threshold of our machining capacity, since this improves the general performance.}
    \label{fig17}
\end{figure}
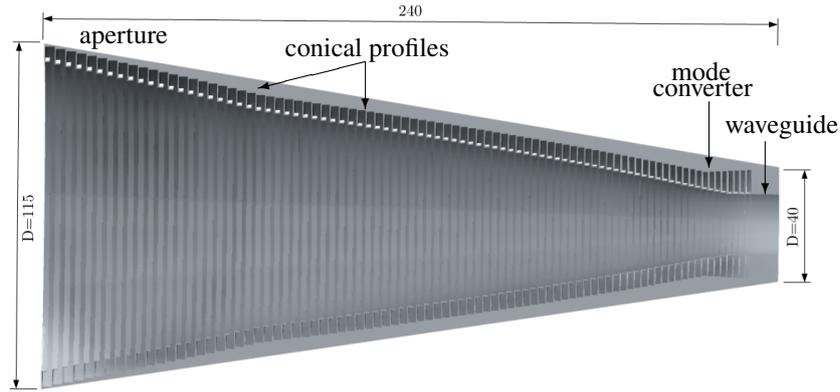

From Eq.~\ref{equation_1}, it follows that a metamaterial satisfies the hybrid-mode condition when $-Z^{S} Z^{P}/Z_0^2$ is exactly 1. We expect that values of $-Z^{S} Z^{P}/Z_0^2$ in the vicinity of 1 yields moderate cross polarization scattering.\footnote{Later we discuss how different features can also influence the levels of cross-polarization.} Therefore, we can empirically define a ``hybrid region'', which is shaded in gray in Fig.~\ref{fig_5}. The solid blue line corresponds to the top-plated or ``inverted L'' metamaterial shown in Fig.~\ref{fig_2}. The acceptable behaviour in terms of cross-polarization is found within the 6.5--35~GHz band, i.e. a 5:1 bandwidth factor. The dashed red line represents the case of a corrugated wall with similar dimensions. Note that the product of normalized impedances has a steep slope, which leads to a bandwidth factor of approximately 1.4:1 or 40\% of the band limited between 17--24~GHz, as expected. The error in the simulations, performed in the time domain, propagates when the results are converted to the frequency domain by Fourier transform. Together with a non-negligible plate thickness compared to the wavelength, of the order of $\lambda/10$, this can explain the observed frequency shift with respect to an idealized model, where a short circuit is expected at the top of the groove around 23~GHz for a depth of 6.5~mm. We found also that it is not possible to improve the performance of the corrugated wall in terms of the hybrid condition. A different geometry results in a shifted band but the slope of $-Z^{S} Z^{P}/Z_0^2$ always remains large, so the \textit{hybrid bandwidth} is always about 40\% or lower. In the next section we present a meta-horn based on the ``inverted L'' meta-ring topology shown in Fig.~\ref{fig_2} (center panel), fabricated in a standard workshop by traditional techniques \cite{DEMIGUELHERNANDEZ2019103195}. Its layout, resulting in a 240~mm long dual-tapered profile with a 103.1~mm aperture diameter, is shown in Fig.~\ref{fig17} which addresses some details about the design.

\section{A broad band meta-horn prototype made of meta-rings}
\label{Fabrication}

\begin{figure}[ht]
\centering
    \includegraphics[height=0.28\textheight]{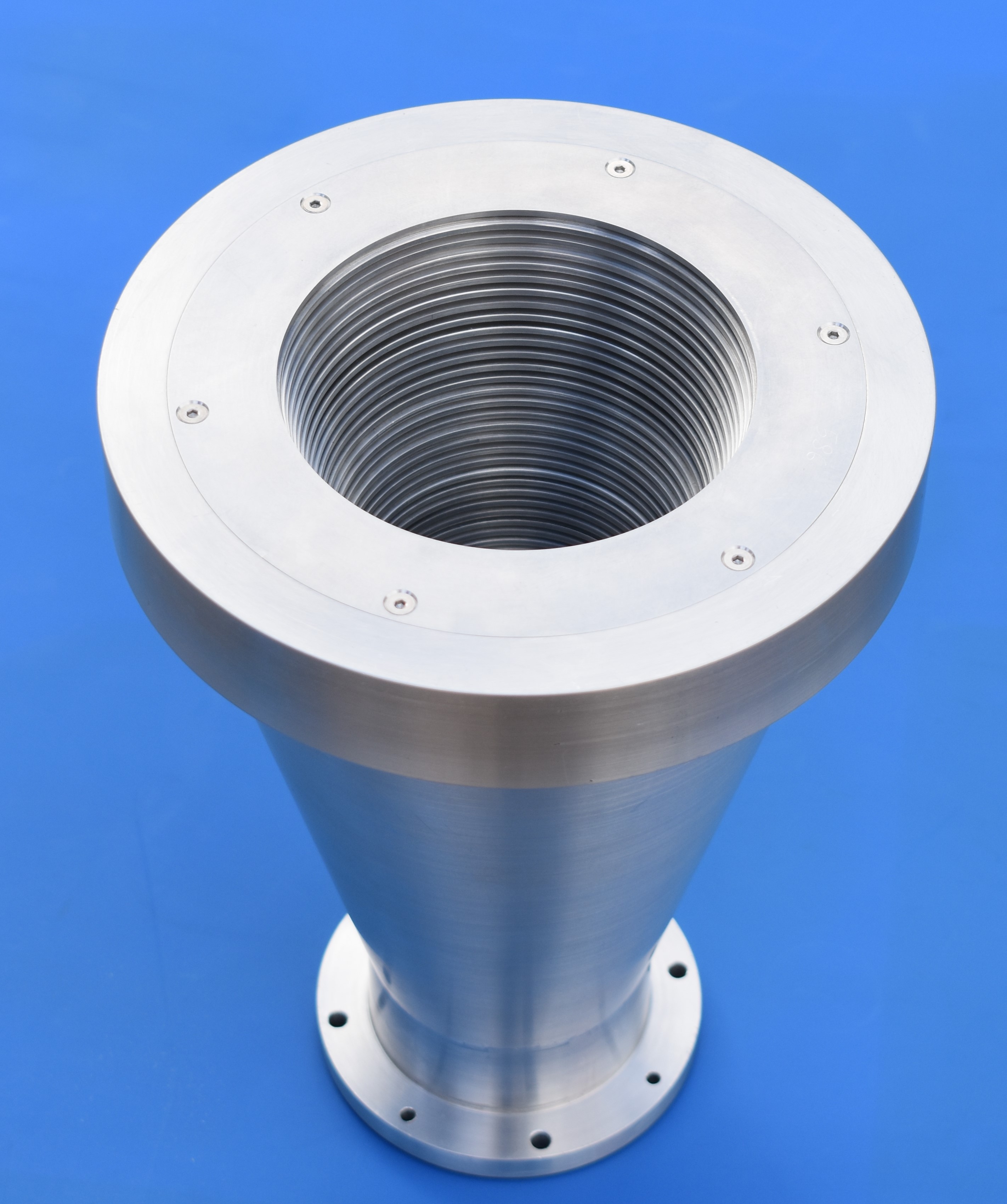}
    \includegraphics[height=0.28\textheight]{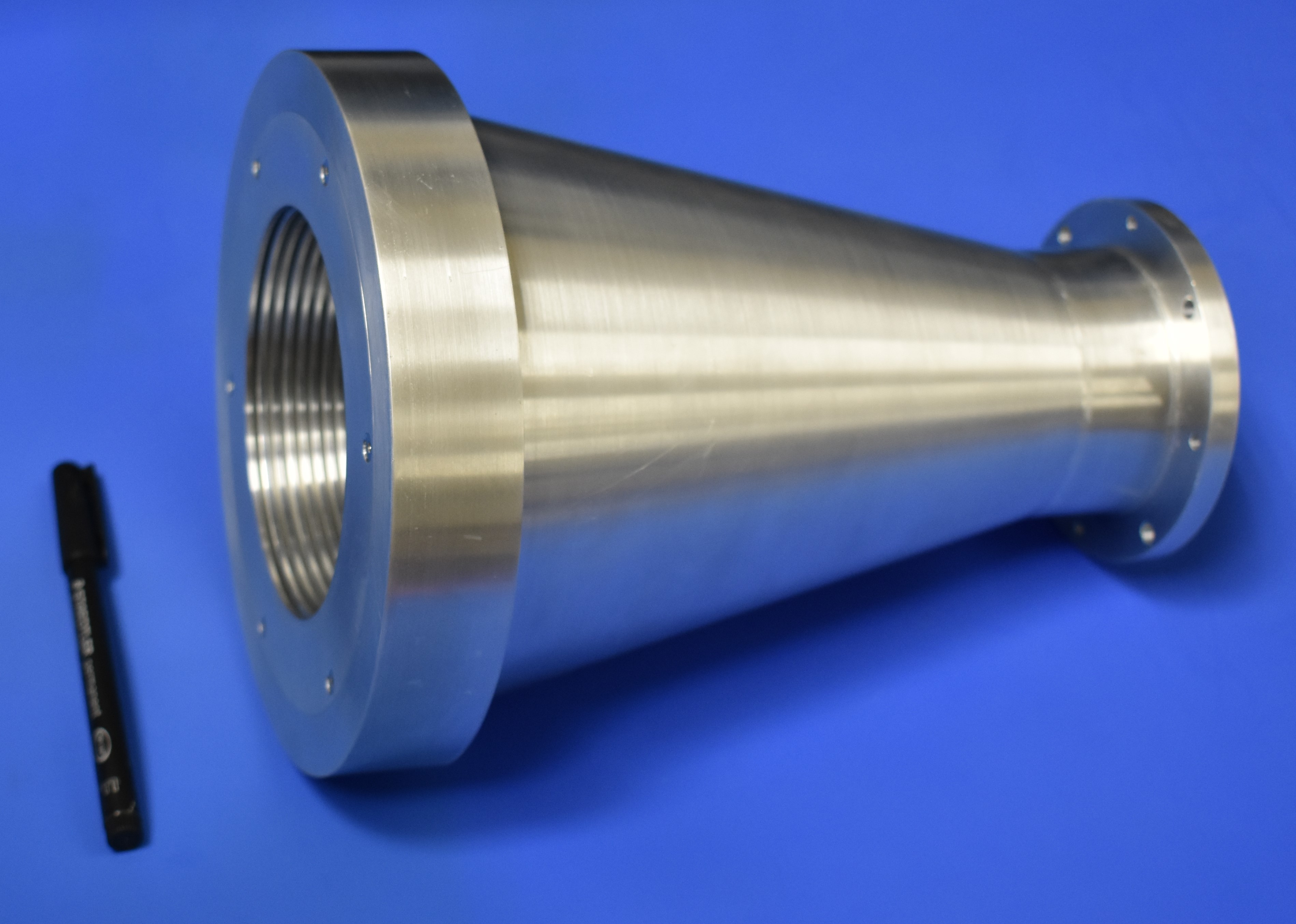}
    \caption{Top (left) and lateral view (right) of the meta-horn prototype. The horn circular waveguide is about 20~mm in diameter and its aperture diameter is 103.1~mm. The horn total length from the base plane to the aperture is 238~mm.}
    \label{fig_6_7}
\end{figure}

Fig.~\ref{fig_6_7} shows the meta-horn prototype, which was manufactured using computer numerical control (CNC) machines.

The 85 meta-rings of the horn were assembled into an Aluminum external shell, which provided stability to the structure. Manufacturing tolerance is satisfactory, with less than 30 $\mu$m of deviation from the nominal value. The metrological analysis was carried out with a coordinate measuring machine following the same methodology described in \cite{DEMIGUELHERNANDEZ2019103195}, i.e. measuring diameter, roundness, thickness and parallelism. A thermally stabilized environment prevented possible deviations due to the expansion or contraction of the materials. By measuring coordinates of one or several points, the machine allowed both dimensional and geometric measurements. Each piece was fixed on the machine bench before making measurements. We had to establish one or more reference system for each piece depending on the dimensions to be verified. In addition, each measurement was processed by multi-point adjustment to average the error. The resolution of the machine is of the order of a few microns.

\begin{figure}
\centering
    \includegraphics[width=0.5\textwidth]{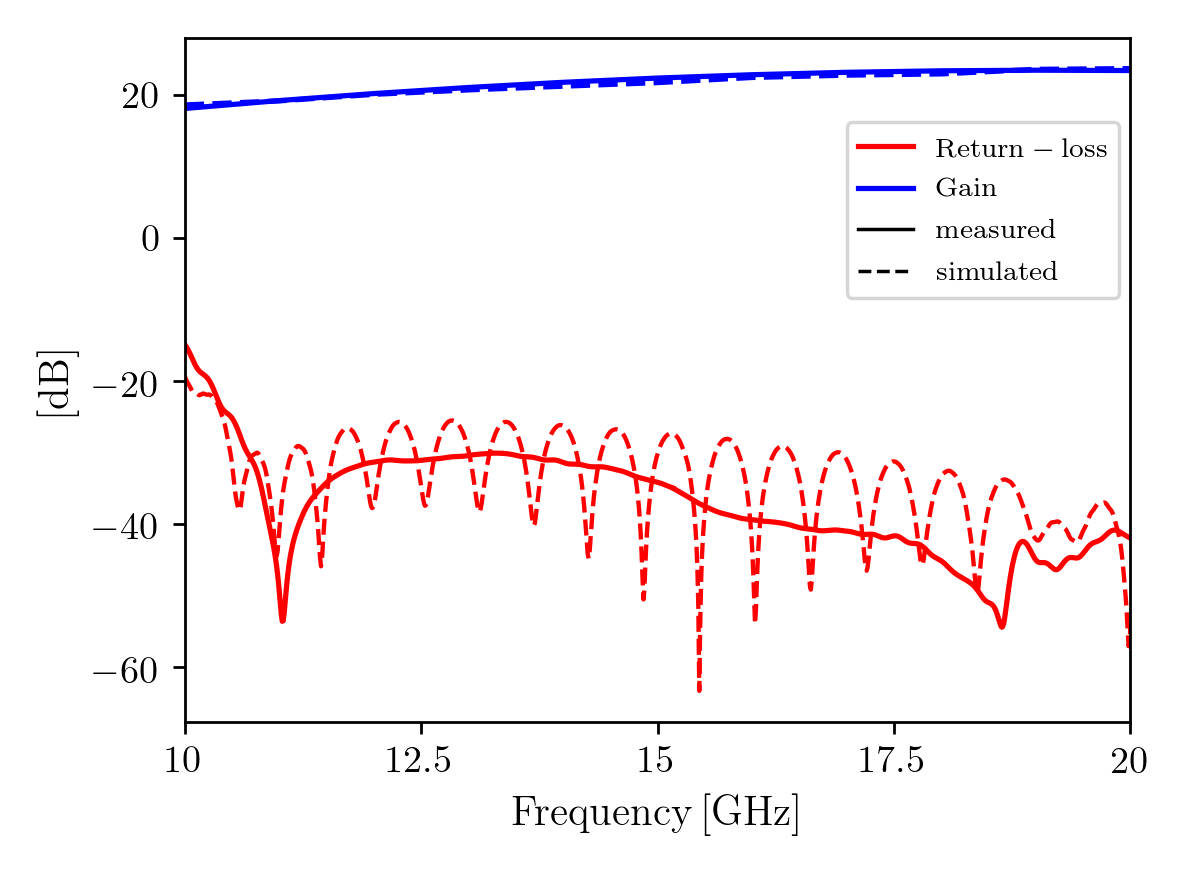}
    \centering 
    \caption{Return-loss (S$_{11}$) and maximal realized gain. Experimental results compared to the result from FEM simulation.}
    \label{fig7_6}
\end{figure}

The electromagnetic measurements were made within an anechoic chamber. For this operation, the 10--20~GHz band was divided into sub-bands defined by the standard commercial bands for microwave components, so all the waveguides operate in a monomode regime. Thus, the experimental set-up includes WR-75 standard gain horns, WR75 FBP120 to SMA-F adapters, and a 20~mm diameter circular-to-rectangular waveguide transition covering the 10--15~GHz range; the 15--20~GHz sub-band is covered by WR-51 standard gain horns, WR51 FBP180 to SMA-F adapters and a 20~mm diameter circular-to-rectangular (FBP180 flange) waveguide transition. Both customized circular-to-rectangular transitions are used to connect the rectangular waveguides of the standard components to the meta-horn, which has a 20~mm diameter input. 
	
\begin{figure}
\centering
    \includegraphics[width=0.7\textwidth]{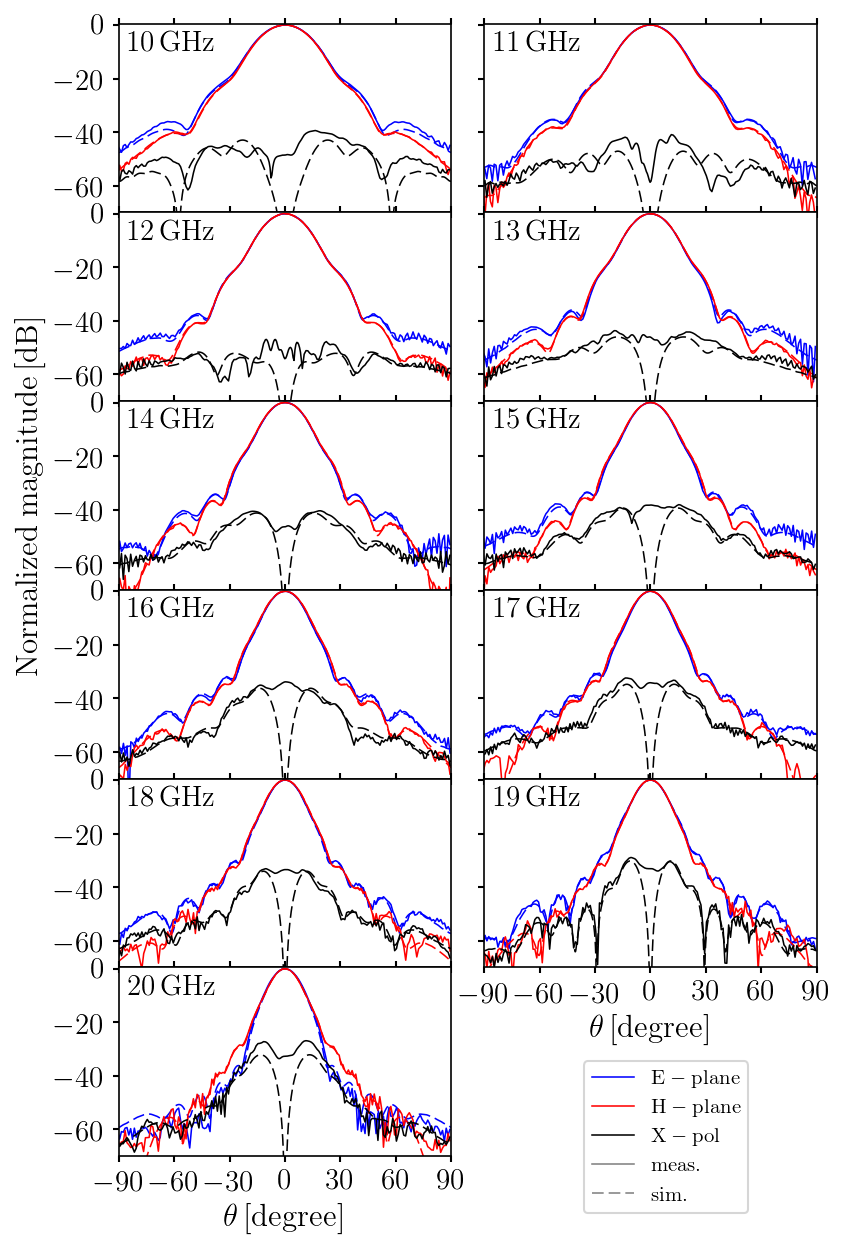}
    \centering 
    \caption{Experimental results compared to Method of Moments (MoM) simulation. The plots show the E- and H-plane antenna patterns and the cross-polarization level (X-pol) from 10 to 20~GHz with 1~GHz steps.}
    \label{fig7_7}
\end{figure}

Radiation patterns and gain are measured in the far field regime. In each sub-band, a standard pyramidal horn antenna acts as a transmitter (Tx) and it is placed about 3660~mm from the antenna under test. The meta-horn far field regime is 1415~mm at its highest frequency (20~GHz), so that the distance between the Tx antenna and the meta-horn during the measurements is at least 2.6 times the far-field distance. The alignment between the Tx antenna and the meta-horn when mounted on their mechanical support structures is performed with a laser and it is verified by means of preliminary radio frequency (RF) measurements. The polarization alignment is performed by minimizing the cross-polar signal when the two antennas are aligned along their boresight. No significant dynamic range limitations exist thanks to the use of a stable external source directly connected to the Tx antenna, thus preventing power losses through long cables to the Vector Network Analyzer (VNA). The receiver is based on a mixer-diplexer chain connected to the SMA adapter, easily accessible on the back side of the polarization rotary table.

The phase of the signal produced by the external generator is unknown to the VNA, so it is necessary to have a reference (fixed) antenna for phase measurement and power stability verification. This is the reason why this set-up requires two standard gain horns for each sub-band. This set-up guarantees optimal dynamics with respect to standard S$_{21}$ measurements at the VNA ports. As a drawback, it prevents the usage of a real-time time-domain filtering; however, the anechoic environment guarantees negligible spurious reflections during the measurements.

The meta-horn gain is measured with a gain transfer technique, i.e. by replacing the antenna under test with a calibrated gain antenna and comparing their output. The effect of waveguide transitions and cable adapter is also de-embedded by measuring their frequency response with dedicated VNA measurements. This is repeated for both 10--15 and 15--20~GHz sub-ranges.

Finally, the return-loss is measured with the antenna mounted on its mechanical support structure while pointing towards the anechoic chamber walls to minimize any spurious contribution from the environment. A time-domain filtering is also applied to verify possible reflections due to the waveguide and cable adapters, as well as possible mechanical misalignments at the of the RF chain interfaces.%\newline

The simulated performance of the meta-horn show a return-loss better than $20$~dB, side-lobe level better than $-30$~dB, cross-polarization level better than $-35$~dB and a 20~dBi gain feature in the 10--20~GHz band, approximately. These performance have been verified in the bandwidth by measurements, as discussed here below.

Figure~\ref{fig7_6} shows a comparison between measurements (solid line) and simulations (dashed line) for the meta-horn return-loss (red) and gain (blue), resulting in a excellent agreement and consistency over the 10--20~GHz frequency band. Fig.~\ref{fig7_7} reports the radiation patterns as measured between 10 and 20~GHz with 1~GHz step. Each plot includes the co-polar E- and H-planes and the cross-polar 45°-plane as compared to the relevant simulation. The agreement between measurements and simulation is at the level of fractions of a dB down to the first side-lobe at all frequencies. Moreover, the cross-polar level predicted by simulations is confirmed by meta-horn measurements. A few dB discrepancy is observed at 11 and 20~GHz,\footnote{We think that such a discrepancy has an instrumental-experimental origin. Although minor, this will be further investigated in the future.} nevertheless guaranteeing a cross-polar maximum level below -40 and -35~dB, respectively.

\begin{figure}
    \centering
    \includegraphics[width=0.5\textwidth]{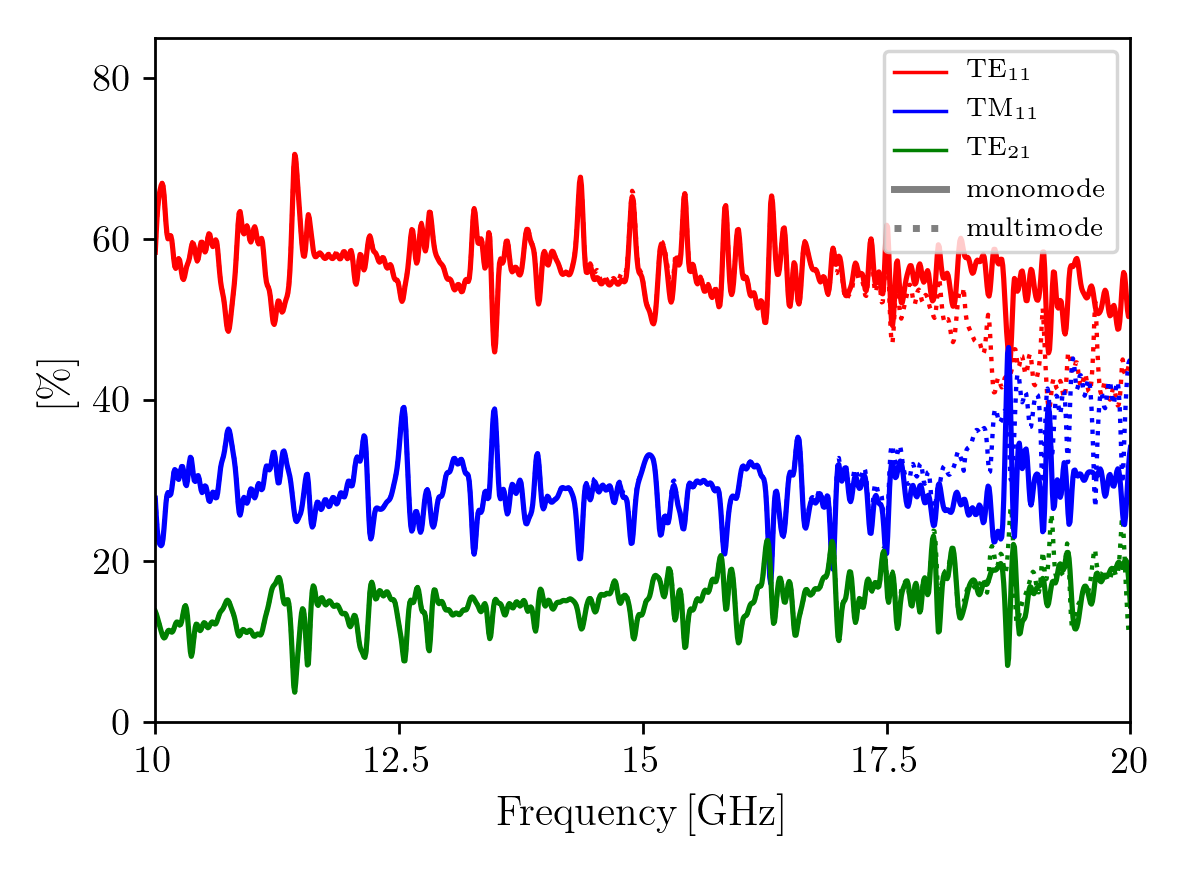}
    \centering 
    \caption{Modal decomposition at the output of the meta-horn. Note the possible effect on TM$_{11}$ transmission due to a multimodal feeding waveguide, shifting around  15\% of the energy between the TE$_{11}$ and TM$_{11}$ modes. The power in other modes is negligible and so is omitted here. The analysis relies on FEM simulation.}
    \label{fig7_8}
\end{figure}	

\begin{figure}
    \centering
    \includegraphics[width=0.5\textwidth]{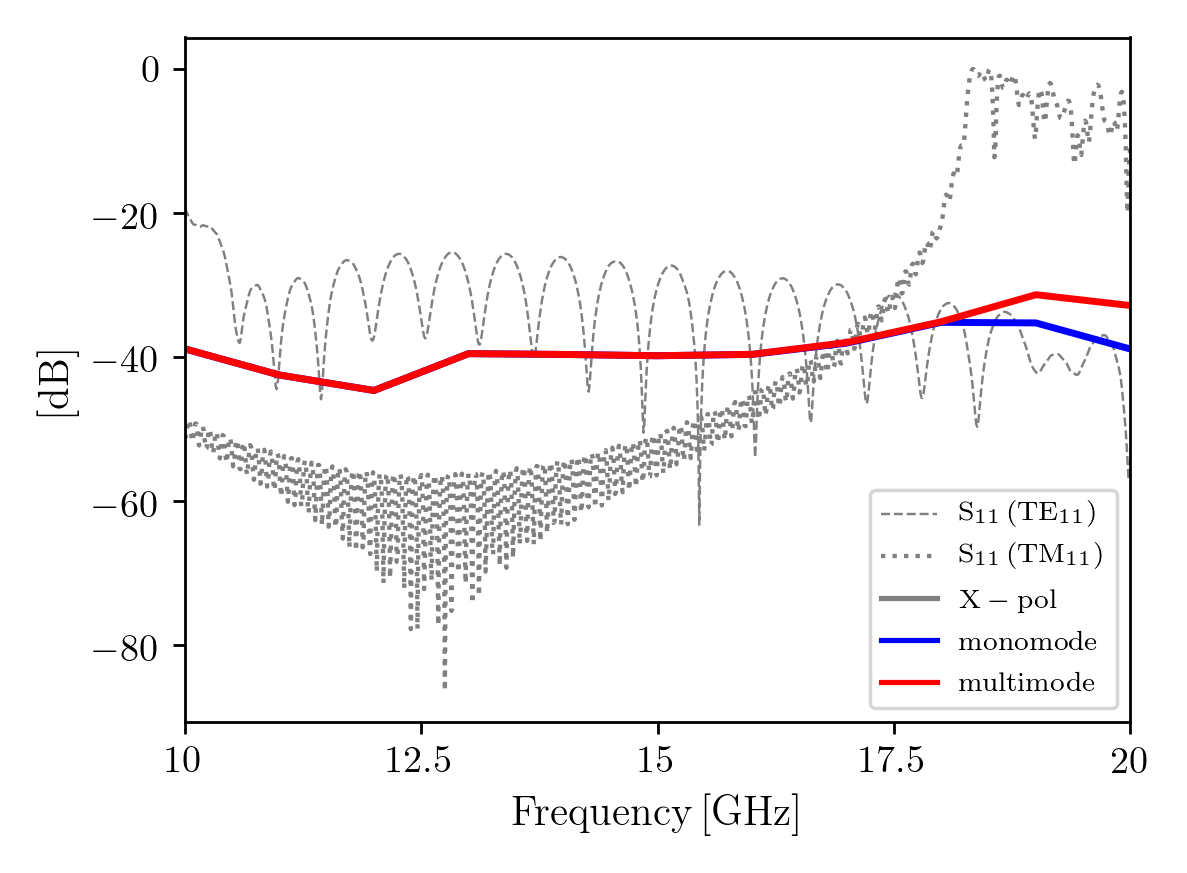}
    \centering 
    \caption{FEM simulations of monomode feeding versus multimode feeding cross-polarization (X-pol). Although the X-pol and return-loss (S$_{11}$) scales are different, note the TM$_{11}$ mode cut-off frequency around 18.3~GHz and its possible contribution to the multimode X-pol figure.}
    \label{fig7_3}
\end{figure}

Theory can help interpret these results. At an early design stage we noticed that, because of the ultra-wide bandwidth of the meta-horn, it was not possible to maintain only the fundamental mode (TE$_{11}$) at the feeding waveguide for a single-mode operation through the complete 2:1 factor band. Simulations results in Fig.~\ref{fig7_8} suggest that around 60\% of the energy is transmitted in the fundamental mode at the output of the meta-horn in the monomodal feed-guide case, and that the remaining power is divided into higher-order modes (mainly TM$_{11}$ and TE$_{21}$), slightly differing from the HE$_{11}$ \textit{hybrid mode} decomposition \cite{HE11} in power. One might think that reducing the radius of the feeding waveguide below a critical size\footnote{For this prototype, a feeding waveguide diameter less than about 9.3~mm raises the cut-off frequency of the mode TM$_{11}$ above 20~GHz.} could be enough for providing an optimal single-mode operation. Unfortunately, a significant feed-guide radius reduction has a negative effect on return-loss that cannot be easily surpassed. On the other hand, simulations with a toy model reveal that optimization with the goal of hybrid-mode condition fulfilment or with the goal of cross-polarization minimization tends to differ slightly, pointing to a higher complexity of the issue from the analytical point of view. This could be explained in the sense that, in a real horn, several factors influence antenna performance, such as the aperture, the length and the profile, as well as the surface impedance as treated throughout this project \cite{Miguel_Hern_ndez_2019}~\cite{Clarricoats}. Furthermore, the modal distribution typically varies through the length (and aperture) of a horn, so different planes (e.g., input and output) can have a different modal decomposition. Obviously, the presence of the TE$_{21}$ mode displaces the presence of other modes with respect to the ideal HE$_{11}$, in whose decomposition this mode is not present. The weight of higher-order modes is even more significant when a multimode feeding waveguide is considered. This is represented in Fig.~\ref{fig7_8} (dashed line), in comparison to the single-mode operation regime, represented by a solid line. The power modal decomposition seems to have an influence on cross-polarization, and a negligible influence on the other figures of merit of the meta-horn. The influence on cross-polarization is represented in Figure~\ref{fig7_3}.

More conclusions and future lines are to be discussed in the following section.

\section{Summary and conclusions}
\label{Results}

In this article we presented a proof-of-concept prototype of an innovative horn antenna based on a novel kind of meta-ring which gives the remarkable advantage of easy manufacture at a low cost. The meta-horn gives a good performance (return-loss better than $20$~dB, side-lobe level better than $-30$~dB, cross-polarization level better than $-35$~dB and gain around 20~dBi) over a broad band, with a 2:1 bandwidth factor (see Figs.~\ref{fig7_6} and \ref{fig7_7}).\footnote{Measured gain represented in Fig.~\ref{fig7_6} results from a second order polynomial curve fit of ten gain measurements between 10--20~GHz in 1~GHz steps.} It is important to emphasize that the device is tunable in frequency by simply scaling its dimensions accordingly, with an upper limit established by the capacity of a workshop during the mechanization of the meta-rings. This would allow one to cover a significant part of the radio frequency spectrum from a few GHz up to hundreds.\footnote{Our simulations reveal that a regular workshop, like the one which manufactured the prototype presented here, is suitable for the production of a meta-horn at frequencies over 200~GHz and probably beyond.}

Finally, the meta-horn presented in this work is the result of a design with reduced complexity. For example, it presents a conical profile and a limited number of meta-rings in order to simplify the fabrication process in a proof-of-principle prototype. Therefore, we believe that it has a significant potential for improvement and a large margin for a further research, incorporating also the know-how acquired during the fabrication and lab verification processes. We found also that the Method of Moments (MoM) is more reliable than FEM in the simulation of the antenna pattern, in contrast with our previous expectations.\footnote{With SRSR-D or Mician $\mu$Wave Wizard compared to CST.} This could help in a future (re-)design stage. 

In addition, The meta-horn presented here was optimized for single-mode operation, since a monomode broad band feed-guide can be incorporated \cite{wbwg1}~\cite{wbwg2}; however, the idea of living with higher order modes and shaping them according to specific purposes is also very attractive for future applications. Given this, a more control over the TM$_{11}$ mode would give an improvement of the meta-horn cross-polarization up to levels under $-40$~dB or less in the multimode regime through a $\geqslant$2:1 bandwidth factor, as shown in Figs.~\ref{fig7_8} and \ref{fig7_3}, while mitigating the side-lobes. We identified several areas to explore the full potential of the meta-horn, such as shaping the horn with a smart curve profile instead of a conical profile (e.g., Gaussian \cite{1173026}) or adding length to the antenna (thus increasing the number of meta-rings) in combination with a further computer aided optimization. This would make it possible to simultaneously cover both the K$_a$ and K$_u$ bands with a single receiver for satellite communications, or to cover broader bands reducing the complexity and enhancing the sensitivity of telescopes for radio astronomy~\cite{8863948}~\cite{DEMIGUELHERNANDEZ2020101367}.

\section*{Acknowledgements}
The authors would like to thank R. Hoyland and M. Bersanelli their support making possible bringing the present work to fruition. Thanks C. del Río for comments.\newline

JDM acknowledges financial support from the Spanish Ministry of Science and Innovation under the FEDER Agreement INSIDE-OOCC (ICTS-2019-03-IAC-12). Also the financial support from projects IACA15-BE-3707 and EQC2018-004918-P, and the Severo Ochoa Program SEV-2015-0548.

\section*{Data release}
The data and a 3D model of the meta-horn are available for non-military use upon request. The antenna model is parametrized, allowing rapid prototyping in a different operating band.

%----------------------------------------
\clearpage
\bibliographystyle{JHEP}
\bibliography{DeMiguel_metahorn}

\end{document}